\documentclass[preprint,12pt,nofootinbib]{revtex4-1}

\usepackage{graphicx}\graphicspath{%
{graph/}}%
\usepackage{amssymb,amsmath,amsfonts}

\begin{document}

\title{Wake Measurements of a Dechirper Jaw\\ with Non-Zero Tilt Angle\footnote[1]{Work supported by the U.S. Department of Energy, Office of Science, Office of Basic Energy Sciences, under Contract No. DE-AC02-76SF00515
} }

\author{Karl Bane, Marc Guetg, and Alberto Lutman}
\affiliation{SLAC National Accelerator Laboratory,Menlo Park, CA 94025}

\begin{center}
\end{center}

\begin{abstract}
The RadiaBeam/SLAC dechirper at the Linac Coherent Light Source (LCLS) is being used as a fast kicker, by inducing transverse wakefields, to {\it e.g.} facilitate Fresh-slice, two-color laser operation. The dechirper jaws are independently adjustable at both ends, and it is difficult to avoid leaving residual (longitudinal) tilt in them during set-up. In this report we develop a model independent method of removing unknown tilt in a jaw.
In addition, for a short uniform bunch passing by a single dechirper plate, we derive an explicit analytical formula for the transverse wake kick as function of average plate offset and tilt angle. We perform wake kick measurements for the different dechirper jaws of the RadiaBeam/SLAC dechirper, and find that the agreement between measurement and theory is excellent.
\end{abstract}

\maketitle

\section*{Introduction}

The RadiaBeam/SLAC dechirper at the LCLS is being used as a fast kicker, to facilitate the Fresh-slice, two-color scheme of generating X-rays~\cite{two_color}. In this mode of operation, after the final linac, the beam is made to pass close by to one jaw of a dechirper module, in order to send the tail of the bunch on a different trajectory than the head on the way to the undulator. 
During alignment of the jaws, each end is moved by an independent motor. Thus, in general, a jaw will tend to have an offset as well as some residual tilt with respect to the beam trajectory.

Typically, for Fresh-slice, two-color operation the dechirper jaw is moved toward the beam and adjusted while observing the size of the effect on the induced, downstream betatron oscillation of the beam. The adjustment is not precise and is done somewhat by feel. There may come a time, however, when it is important to accurately know the location and orientation of the jaw with respect to the beam. In a recent report on wakefield measurements on the dechirper at the LCLS, the agreement between measurement and calculation was found to be excellent, after a slight adjustment to the gap parameter in the theory (in two-plate measurements)~\cite{Guetg, Zemella}. However, because of the possibility of an unknown tilt in the jaws, one could not simply conclude that the discrepancy implied an error in measurement or theory. 

This report uses a model independent method of removing unknown tilt in a jaw of the RadiaBeam/SLAC dechirper. The idea of the method is simple. The average transverse kick (or center of mass kick) experienced by a beam on passing by a dechirper plate depends on a strong inverse power of the offset of beam from plate (minus the third power for short bunches). If we run a procedure that fixes the beam offset at the center of the plate (longitudinally, in $z$) while varying the tilt angle in both positive and negative directions, the average wake kick will trace out a curve that has a minimum at the condition of zero tilt angle. This is precisely the experiment that we have performed and report on here.

In this report we also develop an analytical formula for the wake kick experienced by a short bunch on passing a single dechirper plate, as function of average beam offset and plate tilt angle. This allows us to perform a more precise comparison with measurement than was done before~\cite{Guetg, Zemella}.  

\section*{Theory}

\begin{figure}[htb]
\centering
\includegraphics[draft=false, width=.35\textwidth]{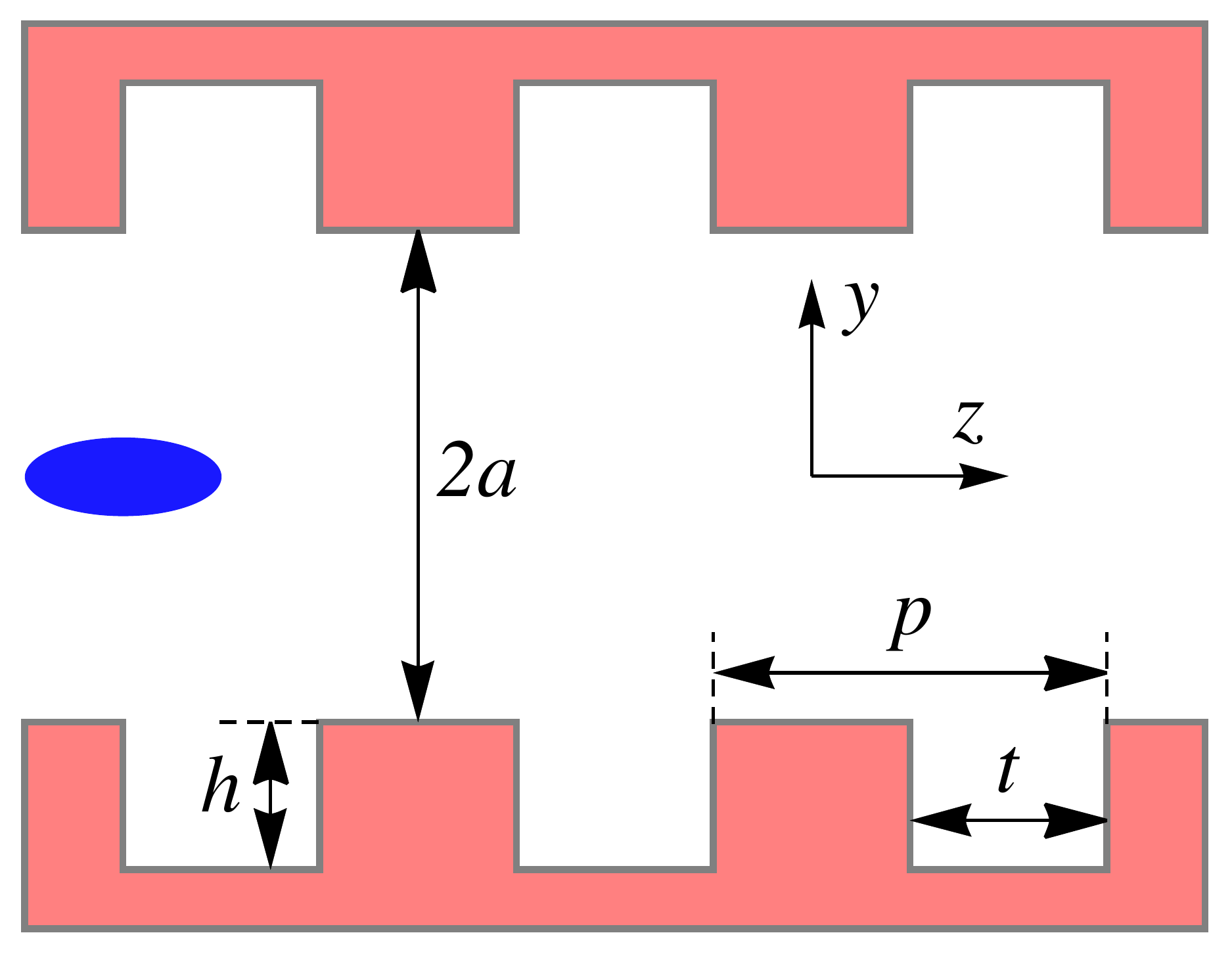}
\caption{Geometry of a (vertical) dechirper showing three corrugations. The blue ellipse represents an electron beam propagating along the $z$ axis.  For the RadiaBeam/SLAC dechirper, (typical) half-gap $a=0.7$~mm, $h=0.5$~mm, $p=0.5$~mm, and $t=0.25$~mm.} \label{geometry_fi}
\end{figure}

The geometry of three corrugations of the RadiaBeam/SLAC dechirper is shown in Fig.~\ref{geometry_fi}. The parameters are (typical) half-gap $a=0.7$~mm, $h=0.5$~mm, $p=0.5$~mm, and $t=0.25$~mm.

\begin{figure}[h]
    \begin{center}
    \includegraphics[width=0.65\textwidth]{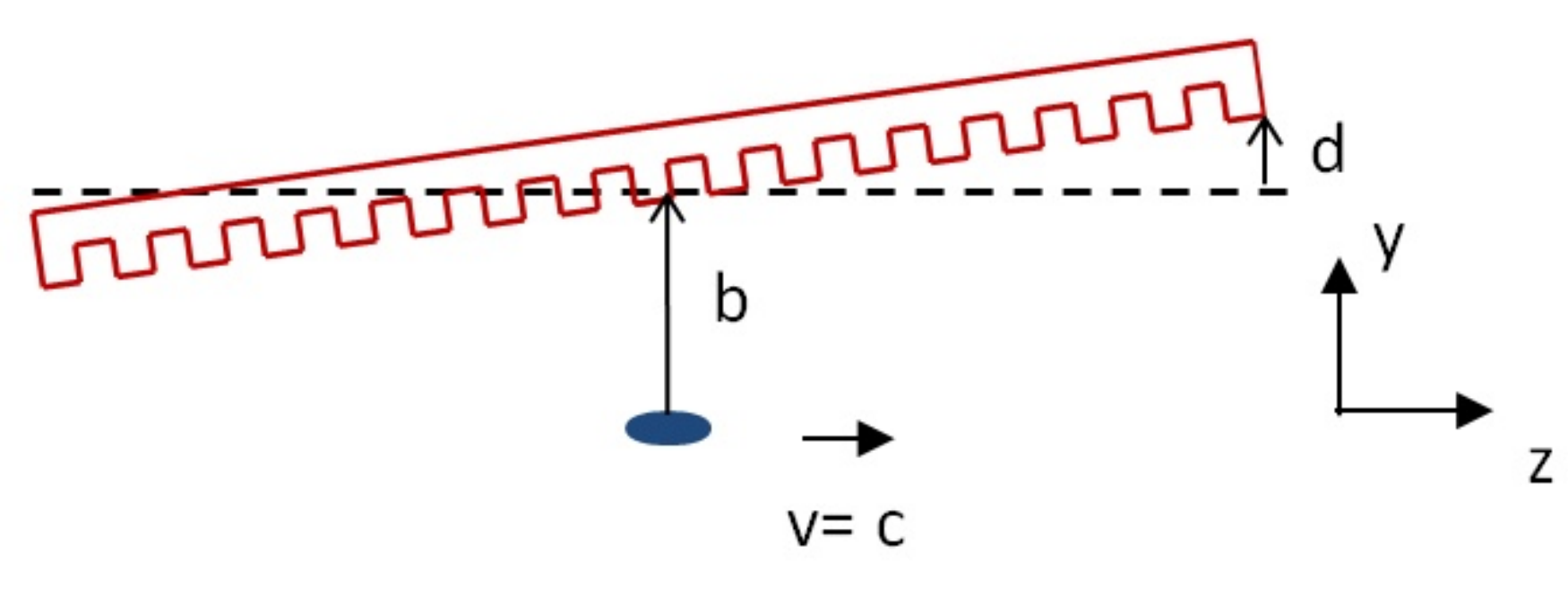}
    \caption{Sketch of orientation of beam and jaw during measurement (for the Top jaw example). The beam (blue ellipse) moves in the $z$ direction below the dechirper jaw (red), at average offset $b$; the jaw tilt (with respect to $z$) is defined by the change in offsets at the jaw ends, $\pm d$. Note that the corrugation size and tilt angle as sketched are much larger than in reality.}\label{sketch_fi}
    \end{center}
    \end{figure}

Let us consider the case of a beam passing by a single dechirper jaw, with the other jaw far away and not interacting with the beam. The ends of the dechirper jaws are independently adjustable. Thus, in general, the configuration of beam and jaw can be characterized by just two parameters, average offset $b$ and extra offset at the jaw ends $\pm d$ (see Fig.~\ref{sketch_fi}; or, equivalently, jaw tilt angle $\tan\theta=2 d/L\approx\theta$, with $L$ the dechirper jaw length). 

 In the measurements to be presented below, the wake strength is quantified by the average transverse kick induced in the beam during its passage near a jaw; this quantity is proportional to the average of the bunch wake, {\it i.e.} the kick factor, $\varkappa_x$.  The bunch at the end of the LCLS linacs is short with an approximately uniform distribution. The kick factor for a short, uniform bunch of full length $\ell$, passing by a single dechirper plate at offset $b$ (with no tilt), is~\cite{Zemella, Bane16, Bane16single}
\begin{equation}
\varkappa_x=\left(\frac{Z_0c}{4\pi b^3}\right)s_{0x}f_x\left(\frac{\ell}{s_{0x}}\right)\ ,\label{kappax_eq}
\end{equation}
with $Z_0=377$~$\Omega$, $c$ the speed of light; with
\begin{equation}
 s_{0x}=\frac{8b^2t}{9\pi \alpha^2 p^2}\ ,
\end{equation}
$\alpha=1-0.465\sqrt{t/p}-0.070(t/p)$, and
\begin{equation}
f_x(\zeta)=1-\frac{12}{\zeta}+\frac{120}{\zeta^2}-8e^{-\sqrt{\zeta}}\left(\frac{1}{\zeta^{1/2}}+\frac{6}{\zeta}+\frac{15}{\zeta^{3/2}}+\frac{15}{\zeta^2}\right)\ .
\end{equation}

For a plate with a small angle tilt, with average offset $b$ and offset at the ends $b\pm  d$, we approximate the total kick factor by averaging the analytical formula along the dechirper plate:
\begin{equation}
\varkappa_{xt}=\frac{1}{2 d}\int_{b- d}^{b+ d} \varkappa_x(b')\,db'\ .
\end{equation}
Substituting in Eq.~\ref{kappax_eq}, and performing the integral we obtain
 \begin{equation}
 \varkappa_{xt}=\left(\frac{Z_0c}{8\pi d}\right)\left[g(b+ d)-g(b- d)\right]\ ,\label{kappa_eq}
 \end{equation}
 with
 \begin{eqnarray}
 g(x)&=&\frac{4t}{9\pi\alpha^2 p^2\xi^2}\Big[6x^2(5x^2-\xi)+e^{-\xi^{1/2}/x}(x\xi^{3/2}-9x^2\xi-30x^3\xi^{1/2}-30x^4)\nonumber\\
 &&+\xi^2\left(Ei[-\xi^{1/2}/x]+\ln x\right)\Big]\ ;
\end{eqnarray}
where
\begin{equation}
\xi=\frac{9\pi\alpha^2 p^2\ell}{8t}\ ,
\end{equation}
and $Ei(x)$ is the exponential integral function. Note that $\varkappa_{xt}$ is symmetric with respect to the variable $ d$, as it should be. 
In Appendix~A we present the corresponding derivation for the longitudinal wake of a tilted plate.

\begin{figure}[h]
    \begin{center}
    \includegraphics[width=0.55\textwidth]{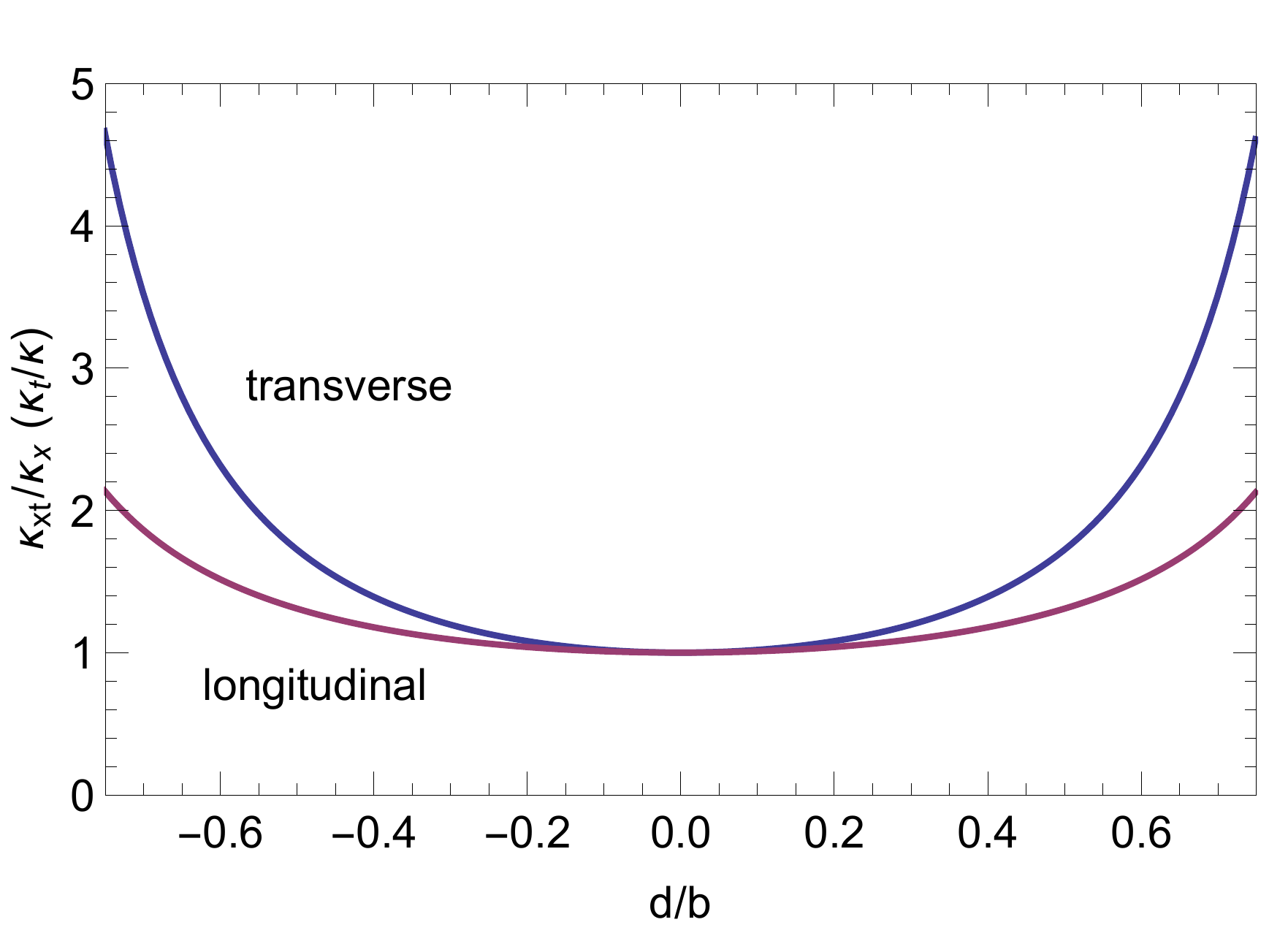}
    \caption{The kick factor for the tilted configuration normalized to the non-tilted case, $\varkappa_{xt}/\varkappa_x$, as function of $ d/b$ (blue). Here we have used as corrugation parameters those of the RadiaBeam/SLAC dechirper, full bunch length $\ell=17$~$\mu$m, and average offset of plate from beam, $b=1$~mm. For the longitudinal case, the change in relative loss factor, $\varkappa_t/\varkappa$, derived in Appendix~A, is given in red.
    }\label{meas_simulation_fi}
    \end{center}
    \end{figure}

The kick factor for the tilted configuration normalized to the non-tilted case, $\varkappa_{xt}/\varkappa_x$, as function of $ d/b$ is shown in Fig.~\ref{meas_simulation_fi} (in blue). Here we have used as corrugation parameters those of the RadiaBeam/SLAC dechirper, full bunch length $\ell=17$~$\mu$m, and average offset of plate from beam, $b=1$~mm. For comparison, the longitudinal effect, {\it i.e.} the change in relative loss factor, $\varkappa_t/\varkappa$, is given in red. We see that the longitudinal wake is a less sensitive function of the tilt than the transverse wake. This is because the longitudinal wake has a weaker dependence on offset of beam from plate, $b$; for a short bunch it varies as $b^{-2}$ instead of the $b^{-3}$ of the transverse case.

The beam position at downstream BPM~590---assuming the beam is initially traveling parallel to the $z$-axis and that there is no intervening magnet---is simply
\begin{equation} 
x_b=eQLL_{BPM}\varkappa_{xt}/E\ ,\label{xb_eq}
\end{equation}
 with $Q$ beam charge, $L$ ($=2$~m) length of dechirper plate, $L_{BPM}$ ($=16.26$~m) distance between dechirper and measuring BPM, and $E$ beam energy. Of course, a usual implicit assumption is that the tail of the bunch does not move significantly,
 compared to offset $b$, during the traversal of the plate.  

\section*{Measurements}

\begin{figure}[h]
    \begin{center}
    \includegraphics[width=0.7\textwidth]{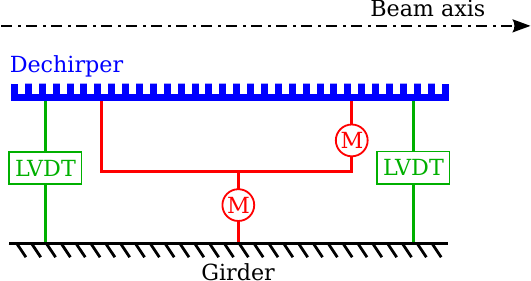}
    \caption{Sketch of the position correction (M stands for ``mover'') and read-back (LVDT sensors) for each dechirper plate. 
     }\label{dechirper_align_fi}
    \end{center}
    \end{figure}

Fig.~\ref{dechirper_align_fi} gives a sketch of the adjustment and read-back system for each dechirper jaw of the RadiaBeam/LCLS dechirper installed in the LCLS. There are two movers  (``M''), and two LVDT (linear variable differential transformer) position sensors, which are located near both ends of the jaw. The main mover (at bottom center of the figure) shifts the entire jaw without changing its inclination (or tilt) angle; the trim mover (the other mover in the figure) moves the downstream end of the jaw. Adjustment of the tilt angle while keeping the average distance to the beam constant, therefore, requires the actuation of both motors. Due to mechanical backlash this can lead to discrepancies between set-points (requested settings) and actual position values. For our measurements, this effect was mitigated by actuating the motors always in one direction, and by using the more accurate LVDT read-back values for data analysis.

For the measurements described below, the beam was kept steady and the dechirper jaws were moved. For each data set one jaw was moved near the beam trajectory (while the three others were moved far away from it), following the sequence the horizontal jaws---North then South---followed by the vertical jaws---Top then Bottom. (It was later discovered that the Top plate was inadvertently left near the beam during the North and South measurements; this, however, should not affect the results presented below.) 
During each measurement the tilt parameter, $ d$, was changed while trying to keep the average offset of the jaw with respect to the beam trajectory, $b$, fixed.
The transverse wake effect was measured using downstream BPM~590.  Simultaneously, the longitudinal wake effect was measured using BPM~693, located in a dispersive region in the beam dump line (unfortunately, unlike in earlier single-plate measurements~\cite{Zemella}, these longitudinal results are extremely noisy, are not much help in the analysis of the measurements, and will not be discussed further.)

\begin{figure}[h]
    \begin{center}
    \includegraphics[width=0.5\textwidth]{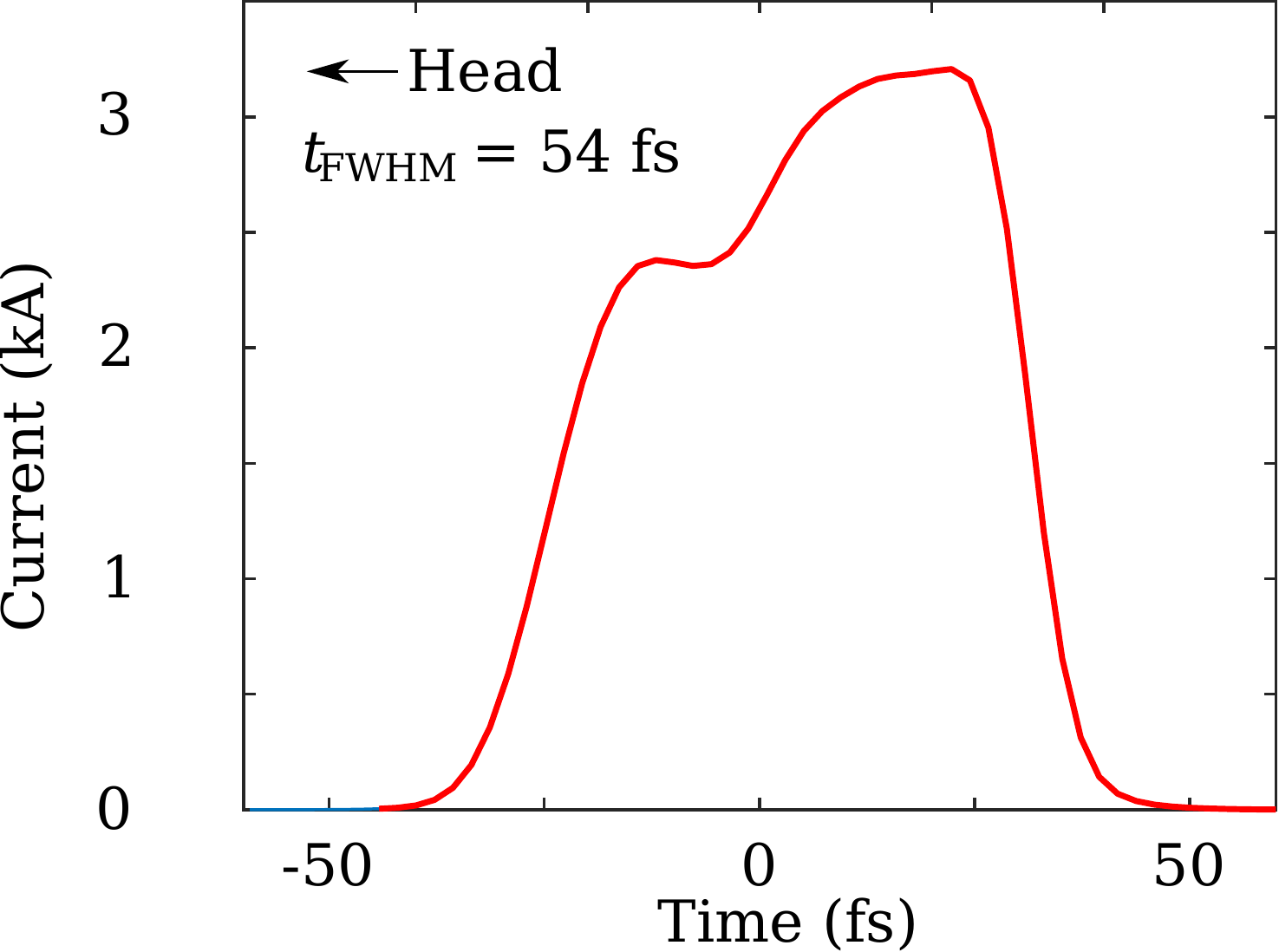}
    \caption{The bunch distribution as obtained by the transverse cavity, XTCAV.
    The head of the bunch is to the left.
    }\label{current_fi}
    \end{center}
    \end{figure}

During the measurements the charge $Q=160$~pC and energy $E=13.24$~GeV. 
 In Fig.~\ref{current_fi} we display the bunch distribution as obtained by the transverse cavity, XTCAV; 
 the head of the bunch is to the left. We see that the distribution is approximately uniform; the uniform distribution with the same area and rms length has peak current $I=2.7$~kA and full length $\ell=18$~$\mu$m.
The transverse beam sizes at the dechirper are $\sigma_x=14$~$\mu$m, $\sigma_y=40$~$\mu$m; our theory assumes that the beam size is small compared to the distance between beam and plate, which is satisfied for our measurements.

\begin{figure}[h]
    \begin{center}
   \includegraphics[width=0.65\textwidth]{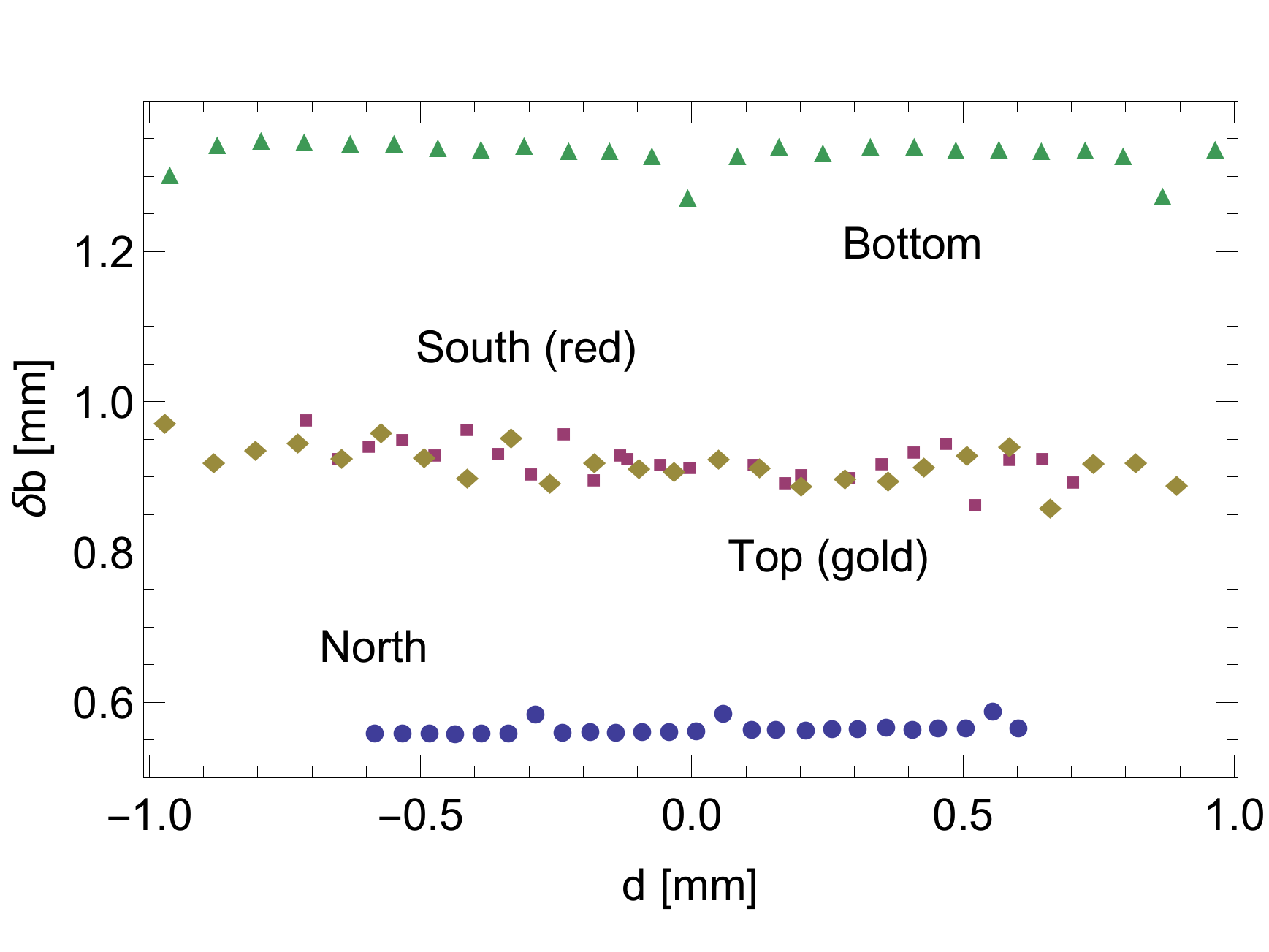}
    \caption{The measured variation in average offset, $\delta b_i$, corresponding to tilt parameter $ d_i$, for all the measurements. The abscissas were shifted by the fitted $ d_0$, to be consistent with the main measurement plots that are given below.
    }\label{deltab_fi}
    \end{center}
    \end{figure}

During data taking, at each stop of the movers, about 100--200 measurements of the BPM~590 reading, $x_b$ (or $y_b$), were obtained. Also, at each stop, the positions of the ends of the plates were recorded by the LVDT sensors. For step $i$, the tilt parameter $d_i$ and average offset $\delta b_i$ are obtained from the LVDT (transverse position) readings $r_{1i}$ and $r_{2i}$ according to
\begin{equation}
d_i=\frac{r_{1i}-r_{2i}}{2}\left(\frac{L}{L-2l}\right)\ ,\quad\quad \delta b_i=\frac{r_{1i}+r_{2i}}{2}\ ,
\end{equation}
where $L$ ($=2$~m) is the dechirper plate length and $l$, the distance between the LVDT sensors and the ends of the plates.
We believe the measured variation in the $\delta b_i$ is fairly accurate, though with a possible, relatively large unknown shift, $b_0$; {\it i.e.} that for the $i^{\rm th}$ measurement point (corresponding to $ d_i$) the actual average offset is
\begin{equation}
b_i=b_0+\delta b_i\ .
\end{equation} 
For each plate, the theory was fit to the data using a nonlinear model fit, with fitting parameters: shift in average offset $b_0$, shift in tilt parameter $ d_0$, and shift in reading on BPM~590, $x_0$ (for North and South) or $y_0$ (for Top and Bottom); the function variables in the fit were $ d$ and $\delta b$. The $\delta b_i$ that were measured are shown in Fig.~\ref{deltab_fi}. We see that the functions were relatively flat during the North and Bottom measurements, but that they had significant variation during the South and Top ones.

\begin{figure}[h]
    \begin{center}
   \includegraphics[width=0.65\textwidth]{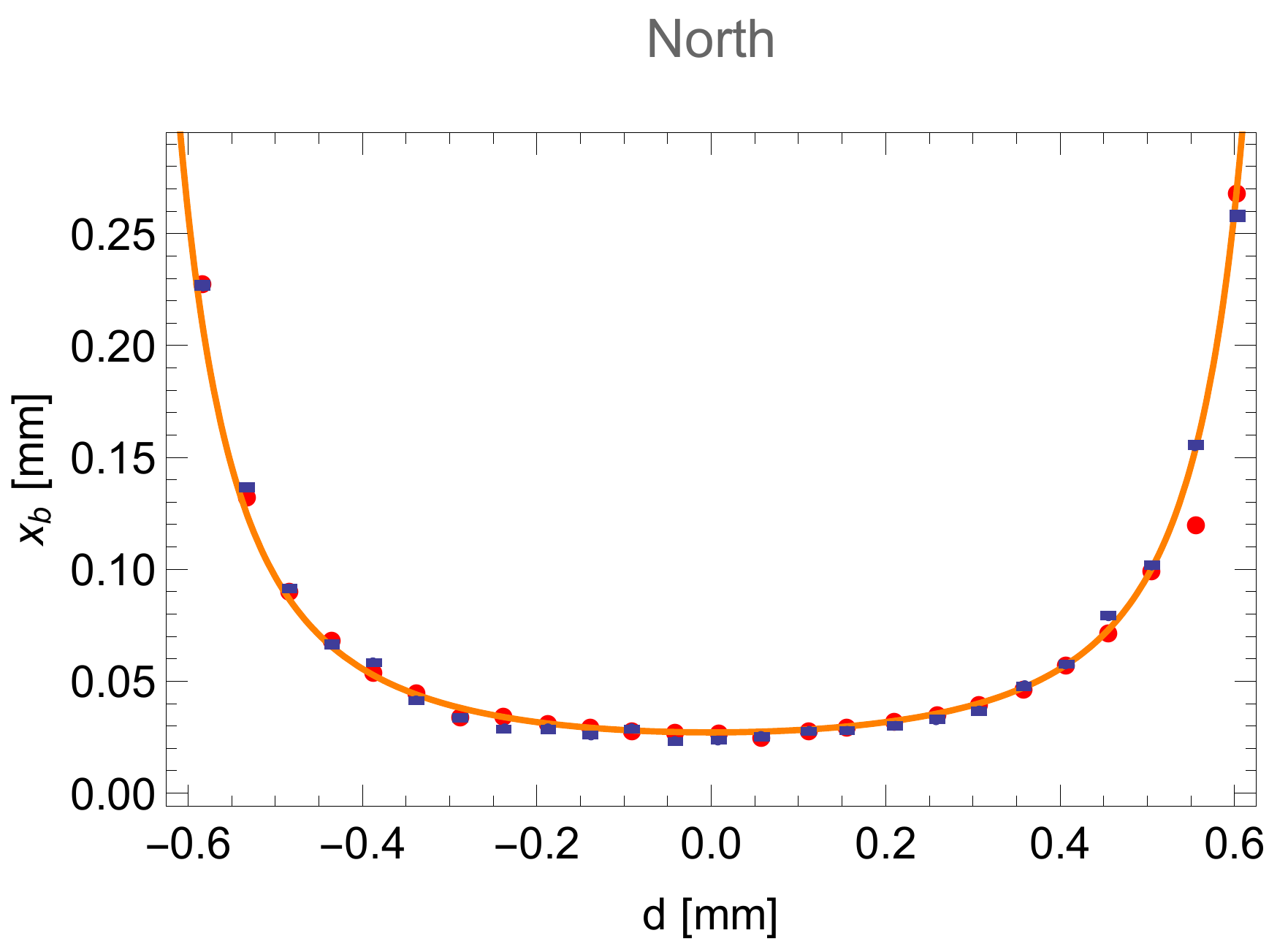}
    \includegraphics[width=0.65\textwidth]{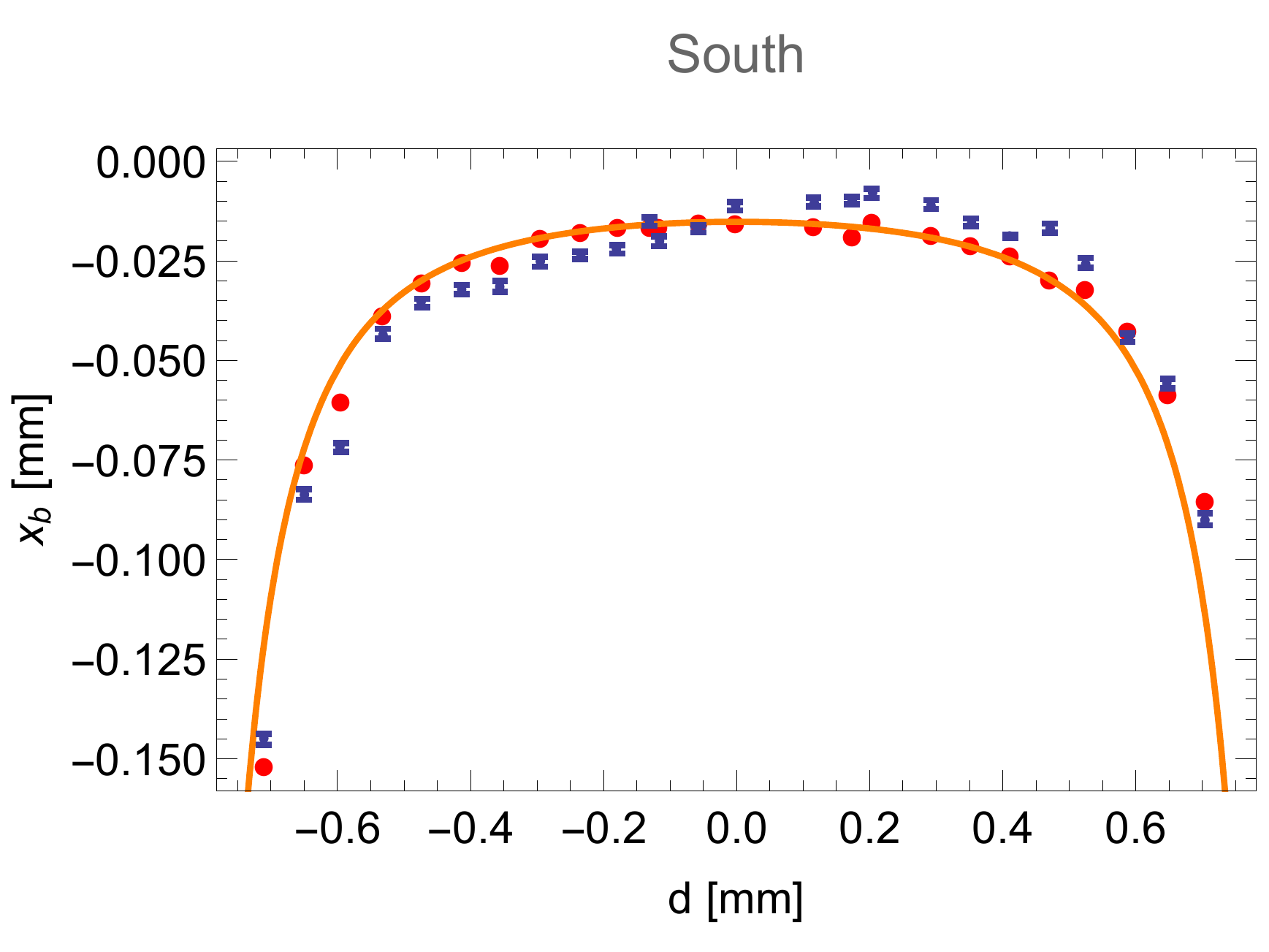}
    \caption{Comparison of measurement (blue plotting symbols) and calculation (orange symbols, orange curve), of beam offset at (downstream) BPM~590, $x_b$, $\it vs.$ jaw tilt parameter, $ d$, for the cases of the North (top plot) and South (bottom plot) dechirper jaws. 
   The blue plotting symbols give the measurement points, after they have been shifted by $-x_{b0}$, $- d_0$, of the fitted parameters; the estimated error in measured $x_b$ (indicated by error bars) is $\sim1.0$ (1.5)~$\mu$m for North (South). The red curve gives the theoretical value assuming $b$ is fixed at $b=\langle b\rangle=0.70$ (0.86)~mm for the North (South) case. The red dots give the best fit of theory to data, when taking as plate offset parameter $b_i=b_0+\delta b_i$ for abscissa value $ d_i$, with $b_0$ the third fitting parameter.
    }\label{ns_fi}
    \end{center}
    \end{figure}

The North results are given in Fig.~\ref{ns_fi} (the top plot), showing the BPM~590 offset, $x_b$, {\it vs.} tilt parameter, $ d$. The fitted parameters are $b_0=0.14$~mm, $x_{b0}=-8$~$\mu$m, $ d_0=-0.42$~mm; the average offset of beam from plate, averaged over all measurement points, was $\langle b\rangle=b_0+\langle \delta b\rangle=0.70$~mm. The blue plotting symbols give the measurement points, after they have been shifted vertically by $-x_{b0}$, horizontally by $- d_0$. For each abscissa value, the rms deviation in measured $x_b$ is $\sigma\sim10$~$\mu$m, and an estimate of the measurement error, $\sigma/\sqrt{N_m}\sim1$~$\mu$m, where $N_m$ is the number of measurements. In this and following figures the data is given with error bars, showing the estimate of measurement error (which however here is tiny and not visible). 
 The red curve gives the theoretical value assuming $b$ is fixed at $b=\langle b\rangle=0.70$~mm. And finally, the red dots give the best fit to the data; {\it i.e.} when taking as average offset $b_i=b_0+\delta b_i$ for abscissa value $ d_i$. A visual measure of the agreement between data and theory is the distance between corresponding blue and red dots.

For the North data, we see that the fit (red dots) to the data (blue dots) is good, and that the local variation in offset $\delta b_i$ has little effect on the results, since almost all the measured points lie on the red curve. 
It is incidentally interesting to note that the initial $ d$ value of the measurements, for which it was assumed that there was little jaw tilt, in fact appeared to have had a significant tilt of $ d= d_0=-0.42$~mm, resulting in an average wake kick that is twice as large as for the case of no tilt.
    
    The corresponding South results are given in Fig.~\ref{ns_fi} (the bottom plot). The estimated error in measured $x_b$ is $\sim1.5$~$\mu$m, and  we see that the error bars have started to become visible. The fitted parameters are $b_0=-0.06$~mm, $x_{b0}=23.$~$\mu$m, $ d_0=-0.15$~mm; $\langle b\rangle=b_0+\langle \delta b\rangle=0.86$~mm.  
We see that the fit (red dots) to the data (blue dots) is good though not as good as before. In addition, from the distance between the red plotting symbols and the red curve we see that the local variation in offset $\delta b_i$ has a small though more significant effect on the results. The right/left asymmetry in the data (the blue dots), which in principle is not allowed by symmetry for this measurement, indicates a systematic error that at the moment is not understood.

\begin{figure}[h]
    \begin{center}
    \includegraphics[width=0.65\textwidth]{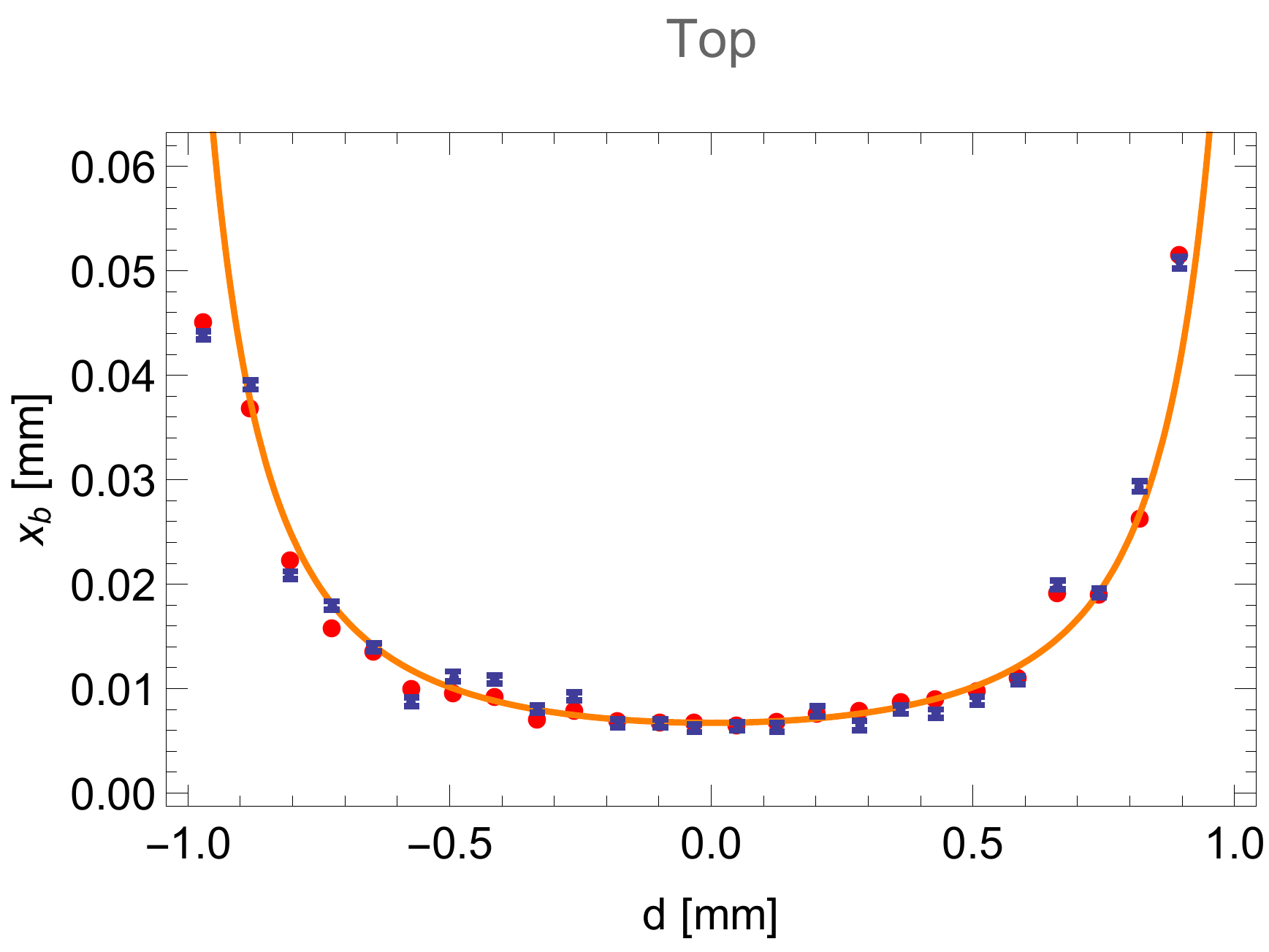}
    \includegraphics[width=0.65\textwidth]{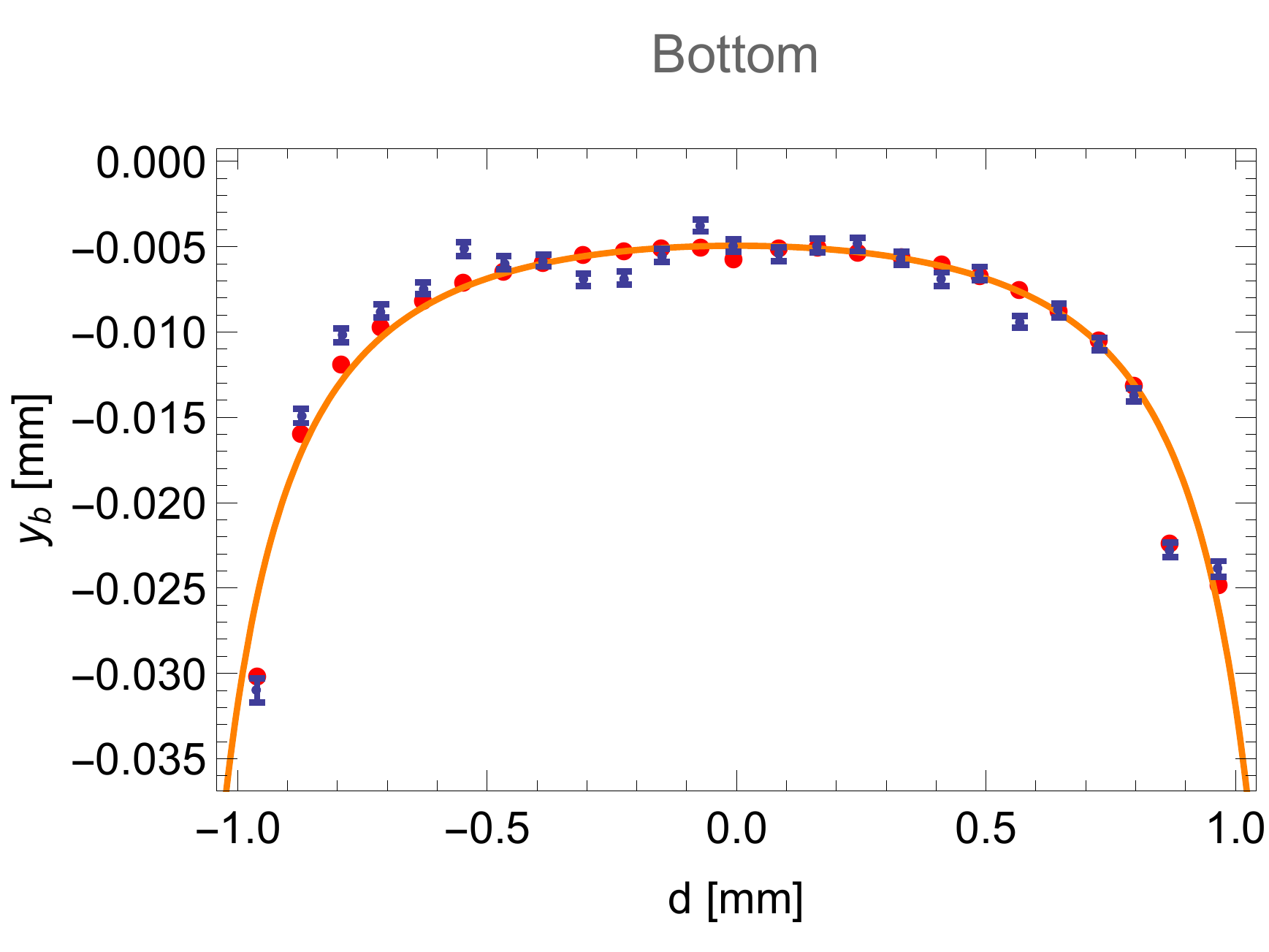}
    \caption{Comparison of measurement (blue plotting symbols) and calculation (orange symbols, orange curve), of beam offset at (downstream) BPM~590, $x_b$, $\it vs.$ jaw tilt parameter, $ d$, for the cases of the Top (top plot) and Bottom (bottom plot) dechirper jaws. 
   The blue plotting symbols give the measurement points, after they have been shifted by $-y_{b0}$, $- d_0$, of the fitted parameters; the estimated error in measured $y_b$ (indicated by error bars) is $\sim0.5$ (0.5)~$\mu$m for Top (Bottom).  The red curve gives the theoretical value assuming $b$ is fixed at $b=\langle b\rangle=1.14$ (1.26)~mm for the Top (Bottom) case. The red dots give the best fit of theory to data, when taking as plate offset parameter $b_i=b_0+\delta b_i$ for abscissa value $ d_i$, with $b_0$ the third fitting parameter.
   }\label{tb_fi}
    \end{center}
    \end{figure}

    The Top and Bottom results are given in Fig.~\ref{tb_fi}. The estimated rms error in measured $y_b$, for both cases, is $\sim5$~$\mu$m. 
    The fitting parameters for the Top case are $b_0=0.22$~mm, $y_{b0}=13.$~$\mu$m, $ d_0=-0.05$~mm; $\langle b\rangle=b_0+\langle \delta b\rangle=1.14$~mm.    
    We see that the agreement with theory is good.
    Note that there are several red-blue pairs of points that agree well though they are off the red curve; this indicates that $\delta b$ has shifted, but nevertheless, theory and measurement are still in good accord.
     For the Bottom case the parameters are $b_0=0.07$~mm, $y_{b0}=1.$~$\mu$m, $ d_0=-0.04$~mm; $\langle b\rangle=b_0+\langle \delta b\rangle=1.26$~mm. 
The agreement between fitted theory and measurement is again good. 
 In Table~I we summarize the fitted results for all four plate measurements. The last column in the table gives the standard deviation of the residuals of the fit to the data, which we see are small in all cases, $10$~$\mu$m or less.

\begin{table}[h!]
\centering
\caption{For the four cases, the fitted: shift in average beam offset, $b_0$; average beam offset from jaw, $\langle b\rangle$; shift in BPM~590 measurement (or equivalently, zero current BPM measurement), $x_{b0}$ or $y_{b0}$; shift in tilt parameter, $ d_0 $; standard deviation of residuals of fit, $\sigma_r$.}
\begin{tabular}{l|c|c|c|c|c} \hline\hline
Case&$b_0$ [mm]&$\langle b\rangle$ [mm]& $x_{b0}$ ($y_{b0}$) [$\mu$m] & $ {d_0}$ [mm] & $\sigma_r$ [$\mu$m]\\ \hline\hline 
North&$0.14$&$0.70$& $-8$&$-0.42$&8\\ \hline
South&$-0.06$&$0.86$& $-23$&$0.15$& 7\\ \hline
Top&0.22&$1.14$& $13$&$0.05$&1\\ \hline
Bottom&0.07&$1.26$& $1$&$$0.04&1\\ \hline
\end{tabular}\label{summary_tab}
\end{table}

From the foregoing results (leaving out those of the South jaw, with their non-physical features), we see that, by performing  single plate wake measurements while sweeping the tilt parameter $ d$, we manage to obtain: the tilt setting that corresponds to zero plate tilt $ d_0$, the shift in average offset parameter $b_0$, and the downstream BPM~590 reading that corresponds to zero wake kick, $x_{b0}$ (or $y_{b0}$). In the previous single plate wake measurements at the LCLS, described in Ref~\cite{Zemella}, the plate was scanned in $b$ while the wake effect was measured at BPM~590. In such a measurement, however, the tilt parameter $ d$ cannot be easily separated from the offset parameter $b$ because of correlations, as explained and demonstrated in Appendix~B. The wake measurement procedure developed in the present report is thus more robust and a significant improvement. 

\section*{Conclusions}

The dechirper jaws of the RadiaBeam/SLAC dechirper at the Linac Coherent Light Source (LCLS) are independently adjustable at both ends, and it is difficult to avoid leaving residual (longitudinal) tilt in them during dechirper set-up and alignment. In this report we present a model-independent method of removing unknown tilt in a dechirper jaw, and demonstrate by experiment that it works well. 

 In addition, we derive an explicit analytical formula for the transverse wake kick of a single dechirper plate, as function of plate offset and tilt angle with respect to the beam orbit. We present wake measurements with the different LCLS dechirper jaws and show that, for the kick factors, agreement with theory is excellent. 
 Compared to previously reported single plate wake measurements that assumed the tilt angle was small and not important~\cite{Zemella}, the measurements reported here are a more sensitive test and stronger confirmation of the theory. 

Having demonstrated the accuracy and sensitivity of this measurement procedure for orienting a dechirper jaw, we propose incorporating it routinely in the set-up and alignment of the RadiaBeam/SLAC dechirper at the LCLS. The procedure is relatively simple and quick to perform.

In Appendix~A we derive an explicit analytical formula for the longitudinal wake kick of a single dechirper plate, as function of plate offset and tilt angle with respect to the beam orbit.

\section*{Appendix A: Longitudinal Effect}

For a uniform bunch distribution, the loss factor---the average of the {\it longitudinal bunch wake}---is given by~\cite{Zemella}
\begin{equation}
\varkappa=\left(\frac{Z_0c}{4\pi b^2}\right)f_z\left(\frac{\ell}{s_{0l}}\right)\ ,\label{kappaz_eq}
\end{equation}
with
\begin{equation}
 s_{0s}=\frac{2b^2t}{\pi \alpha^2 p^2}\ 
\end{equation}
and
\begin{equation}
f_z(\zeta)=\frac{2}{\zeta}\left(1-\frac{6}{\zeta}\right)+e^{-\sqrt{\zeta}}\left[\frac{4}{\zeta}\left(1+\frac{3}{\zeta}\right)+\frac{12}{\zeta^{3/2}}\right]\ .
\end{equation}

For a plate with a small angle tilt, with average offset $b$ and offset at the ends $b\pm  d$, we approximate the total loss factor as
\begin{equation}
\varkappa_{t}=\frac{1}{2 d}\int_{b- d}^{b+ d} \varkappa(b')\,db'\ .
\end{equation}
Substituting in Eq.~\ref{kappaz_eq}, and performing the integral we obtain
 \begin{equation}
 \varkappa_{t}=\left(\frac{Z_0c}{8\pi d}\right)\left[g_z(b+ d)-g_z(b- d)\right]\ ,
 \end{equation}
 with
 \begin{equation}
 g_z(x)=\frac{2x}{\xi_z^2}\left[-2x^2+2x e^{-\xi_z^{1/2}/x}(x+\xi_z^{1/2})+\xi_z\right]\ ,
\end{equation}
where
\begin{equation}
\xi_z=\frac{\pi\alpha^2 p^2\ell}{2t}\ .
\end{equation}

\section*{Appendix B: Comparison of Methods of Single Plate Wake Measurements}

In Ref.~\cite{Zemella} single plate wake measurements were performed and compared with theory. There the plate was assumed to have negligible tilt and the transverse wake kick was measured as function of plate offset from the beam, $b$. Simultaneously, the longitudinal wake effect was also measured as function of $b$. As here, it was assumed that the measured offsets $b$ contained an unknown overall shift $b_0$ that could be significantly larger than the relative error between two $b$ settings. The measurement was performed for the North and the South jaw. The fitted shift values, transverse (longitudinal) were $b_0=-160$~$\mu$m ($-140$~$\mu$m) for the North jaw, and $b_0=-60$~$\mu$m ($-70$~$\mu$m) for the South jaw. The fact that the fitted $b_0$ agreed to within $20$~$\mu$m, for both jaw measurements, gave the authors confidence that the jaw tilt for these measurements was indeed small, and that the measurements confirmed the theory.

\begin{figure}[ht]
    \begin{center}
    \includegraphics[width=0.60\textwidth]{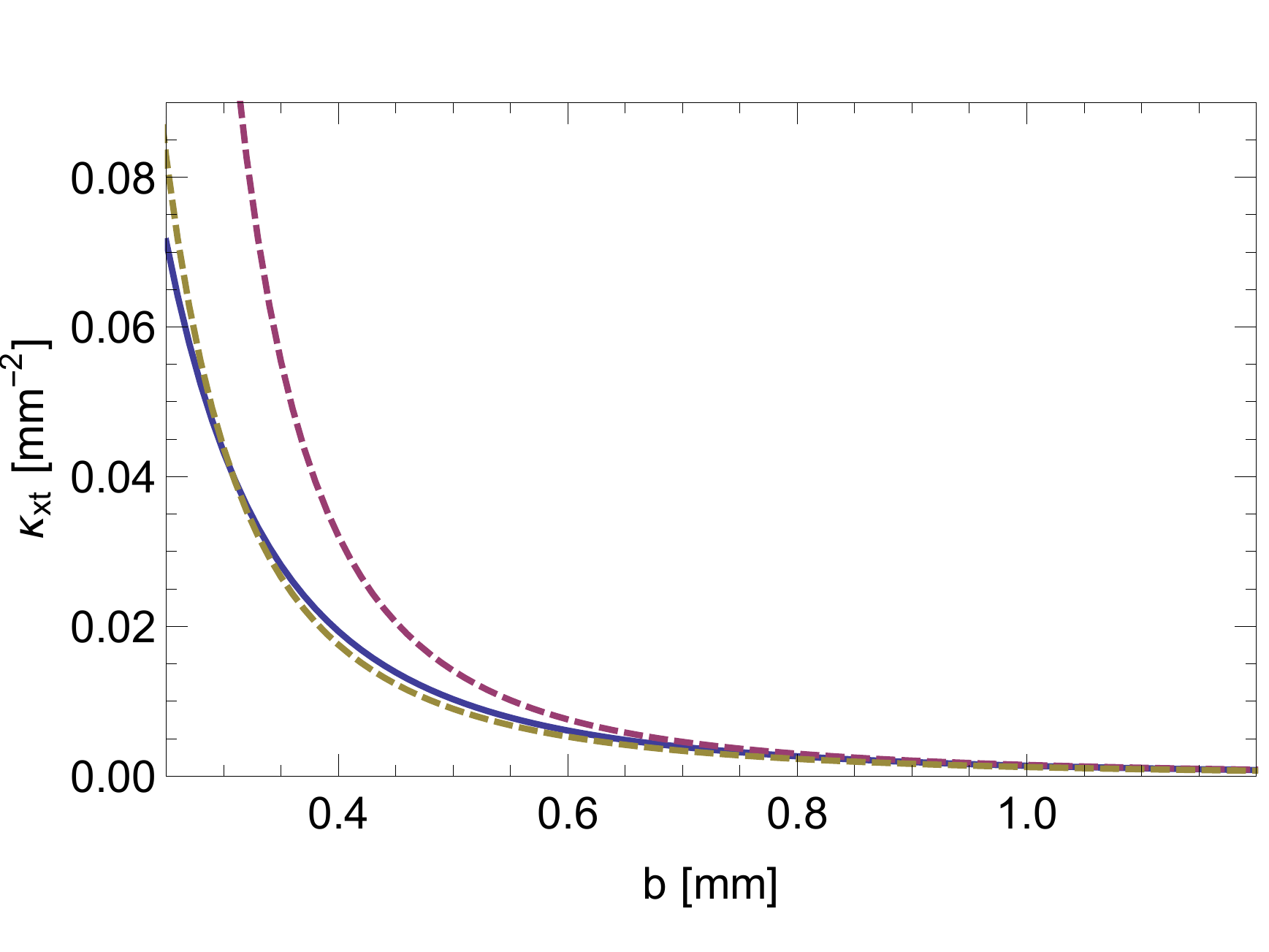}
    \caption{The calculated kick factor $\varkappa_{xt}$ as function of beam offset $b$ assuming no tilt in the jaw (blue solid curve). For comparison, the results for tilt parameter $ d=0.2$~mm are given in red dashes; those for tilt parameter $ d=0.2$~mm and shift by offset parameter $b_0=-70$~$\mu$m are given in gold dashes.
     The corrugation parameters used were those of the RadiaBeam/SLAC dechirper; the bunch distribution assumed was uniform of full length $\ell=18$~$\mu$m.
   }\label{appendix1_fi}
    \end{center}
    \end{figure}

However, as a stand-alone measurement of the transverse wake of a single plate, the measurement described in the present report is much superior to that of Ref.~\cite{Zemella}. To see why, consider Fig.~\ref{appendix1_fi},  which simulates the earlier type of measurement. The corrugation parameters used were those of the RadiaBeam/SLAC dechirper; the bunch distribution assumed was uniform of full length $\ell=18$~$\mu$m. Fig.~\ref{appendix1_fi} shows the simulated kick factor $\varkappa_{xt}$ as function of beam offset $b$ assuming no tilt in the jaw (the blue solid curve).  On the same plot we present results for tilt parameter $ d=0.2$~mm (red dashes), which differ significantly from the blue curve. However, when we shift the abscissa of these last results by $b_0=-70$~$\mu$m, we obtain the curve of the gold dashes, which is now close to the blue curve. 
Thus, there is significant correlation between the parameters $d$ and $b_0$, and using them as fitting parameters for such a measurement will not reliably find their separate values.
Consequently, this kind of measurement is not a good way to find the offset and tilt of a dechirper with respect to the beam.

\begin{figure}[ht]
    \begin{center}
    \includegraphics[width=0.60\textwidth]{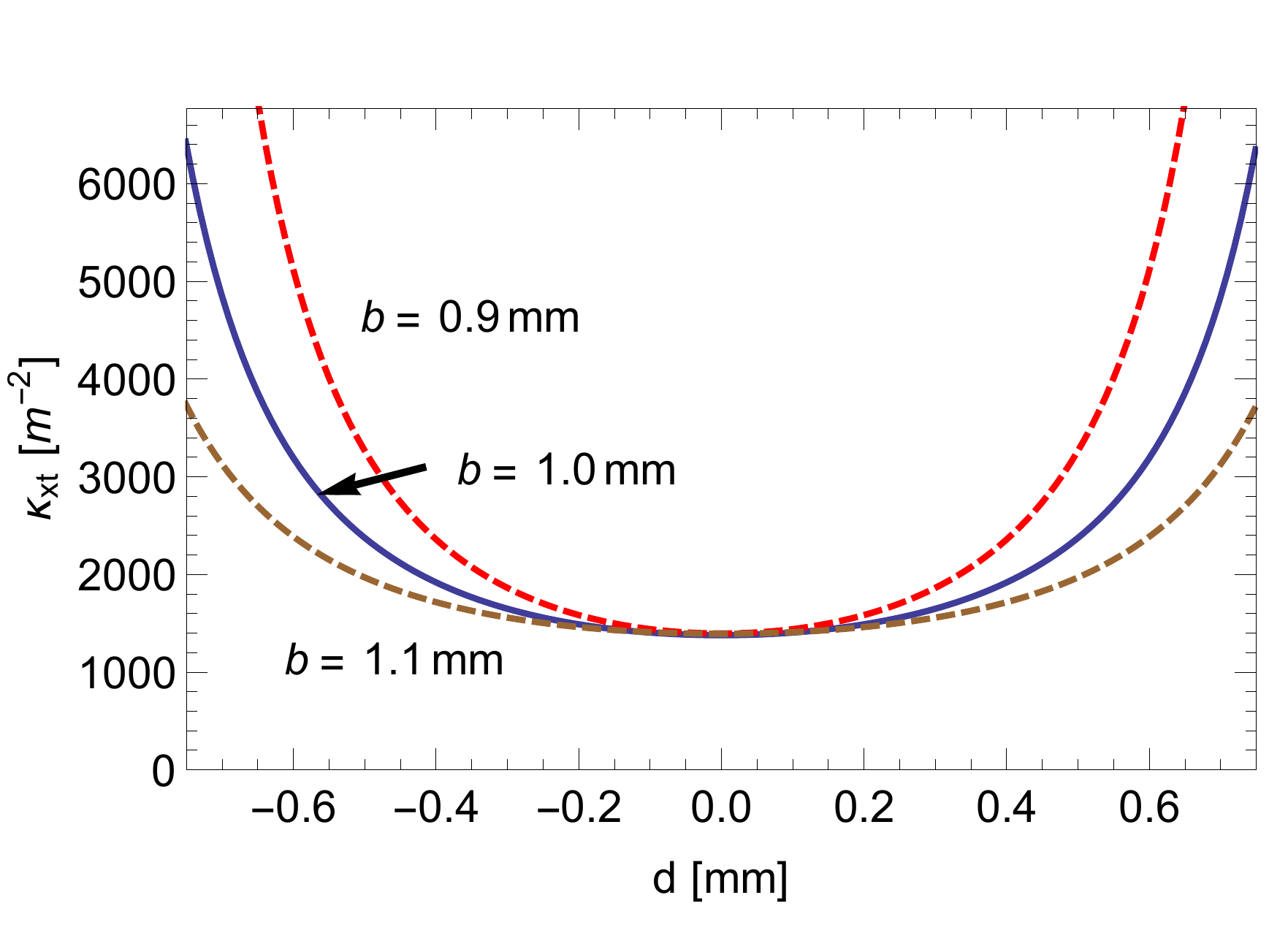}
    \caption{The calculated kick factor $\varkappa_{xt}$ as function of tilt parameter $ d$ for offset parameter $b=1.0$~mm (blue solid curve). For comparison, the results for $b=0.9$~mm (red dashes) and $b=1.1$~mm (brown dashes) are also shown. To emphasize the difference in curvature of the three curves, the latter two were shifted vertically (by, respectively, $-480$~m$^{-2}$, $+350$~m$^{-2}$) to give the same minimum.
     The corrugation parameters used were those of the RadiaBeam/SLAC dechirper; the bunch distribution assumed was uniform of full length $\ell=18$~$\mu$m.
   }\label{sensitivity_fi}
    \end{center}
    \end{figure}

In contrast, consider Fig.~\ref{sensitivity_fi}, which is a simulation of the type of measurement described in the present report. We plot $\varkappa_{kt}$ {\it vs.} tilt parameter $ d$ for cases average offset $b=0.9$~mm (red dashes), 1.0~mm (blue solid line), and 1.1~mm (brown dashes). The dashed curves have been shifted vertically so that all have the same minimum value. One can easily see that there is little correlation between average offset $b$ and a vertical shift in $\varkappa_{xt}$, and that this measurement is very sensitive to changes in $b$. We can further conclude that the measurements for the North, Top, and Bottom dechirper plates (Fig.~\ref{ns_fi}, top plot; Fig.~\ref{tb_fi})~\cite{statement} indicate that the formula for the kick factor of a single corrugated plate, as functions of beam offset and plate tilt (Eq.~\ref{kappa_eq}), is correct to very good accuracy. 

\begin{thebibliography}{1}
\bibitem{statement}
The South plate measurement (Fig.~\ref{ns_fi}, bottom plot) is not included in consideration here, because of nonphysical, asymmetric features in the data that indicate systematic errors. The source of these errors is presently not understood.
\end{thebibliography}
\end{document}